\documentstyle[12pt]{article}

\begin{document}
\hfill
\vbox{
\hbox{CCUTH-96-03} 
\hbox{NHCU-HEP-96-15}}
\setlength{\textwidth}{5.0in}
\setlength{\textheight}{7.5in}
\setlength{\parskip}{0.0in}
\setlength{\baselineskip}{18.2pt}
\vfill
\centerline{\large{{\bf Three-Scale Factorization Theorem and}}}\par
\centerline{\large{{\bf Effective Field Theory}}}\par
\vskip 1.0cm
\centerline{Chia-Hung V. Chang$^1$ and Hsiang-nan Li$^2$}
\vskip 0.3cm
\centerline{$^1$Department of Physics, National Tsing-Hua University,}
\centerline{Hsin-Chu, Taiwan, R.O.C.}
\vskip 0.3cm
\centerline{$^2$Department of Physics, National Chung-Cheng University,}
\centerline{Chia-Yi, Taiwan, R.O.C.}
\vskip 1.0cm
\centerline{\today }
\vskip 2.0 cm
\centerline{\bf Abstract}

We develop a perturbative QCD factorization theorem which is compatible 
with effective field theory. The factorization involves three scales: an 
infrared cutoff of order $\Lambda_{\rm QCD}$, a hard scale of order the $B$ 
meson mass, and an ultraviolet cutoff of order the $W$ boson mass. 
Our approach is renormalization group invariant and moderates the 
scale-dependence problem in effective field theory. 
Applying this formalism to exclusive nonleptonic $B$ meson decays, we
clarify the controversy over the Bauer-Stech-Wirbel parameters
$a_2/a_1$ for charm and bottom decays. 
It is found that the nonfactorizable contribution plays an important
role in the explanation of the sign and magnitude of $a_2/a_1$.


\vfill

\newpage

Nonleptonic heavy meson decays are difficult to analyze due to the 
complicated QCD corrections. While the semileptonic decays involve only 
conserved currents, the nonleptonic decays are described by four-quark 
current-current operators, which is part of the low-energy effective 
Hamiltonians for the W boson exchange. 
For example, the relevant operator for the $B\to D\pi$ decays is 
\begin{equation}
H= \frac{4G_{F}}{\sqrt{2}} V_{cb} V^{\ast}_{ud} 
   (\bar{c}_{L}\gamma_{\mu} b_{L})
            (\bar{d}_{L}\gamma^{\mu} u_{L}) \;,
\label{FourFer}
\end{equation}
The QCD corrections will generate operator mixing among these operators.
The mixing, characterized by Wilson coefficients, depends on an arbitrary 
renormalization scale $\mu$. 
The full effective Hamiltonian related to Eq.~(\ref{FourFer}) can be 
written as 
\begin{equation}
H_{\rm eff} = \frac{4G_{F}}{\sqrt{2}} V_{cb} V^{\ast}_{ud} 
          [ \, c_{1}(\mu) O_{1} + 
            c_{2}(\mu) O_{2} \, ]\;,
\label{eff}
\end{equation}
with
\begin{equation}
O_{1}  =  (\bar{c}_{L}\gamma_{\mu} b_{L})
            (\bar{d}_{L}\gamma^{\mu} u_{L})\;,  \hspace{5ex} 
O_{2}  =  (\bar{d}_{L}\gamma_{\mu} b_{L})
            (\bar{c}_{L}\gamma^{\mu} u_{L}) \;.
\end{equation}
In Eq.~(\ref{eff}) $c_1$ and $c_2$ are the Wilson coefficients, 
whose evolution from the $W$ boson mass $M_W$ down to a lower scale
is determined by renormalization-group running \cite{Buras}. 
Though the Wilson coefficients are $\mu$ dependent, physical quantities 
such as decay amplitudes are not. In principle, the matrix elements of the 
four-fermion operators contain a $\mu$ dependence, which exactly cancels 
that of the  Wilson coefficients. In practical applications, however, 
various schemes are needed to estimate the hadronic matrix elements, and
the estimates are usually $\mu$ independnt. Hence, the decay amplitudes 
turn out to be scale dependent. Take exclusive nonleptonic $B$ meson 
decays as an example, to which the conventional approach is the 
Bauer-Stech-Wirbel (BSW) factorization approximation \cite{BSW}. It is 
assumed that nonleptonic matrix elements can be factorized into two matrix 
elements of (axial) vector currents. Since the currents are conserved, the 
matrix elements have no anomalous scale dependence. 
Presumably $\mu$ should be set to the dominant scale of the matrix 
elements. However, the matrix elements involve both the heavy quark 
scale and the small hadronic scale.
Naively setting $\mu$ to the heavy quark mass will 
lose large logarithms associated with the hadronic scale. 
It is then quite natural that theoretical predictions are sensitive to the 
scale we choose \cite{NRSX,LSW}. 

To circumvent this problem, a phenomenological approach is adopted to 
bypass the strong scale dependence. The Wilson coefficients $c_i$ are
regarded as free parameters, and are determined by experimental data 
\cite{BSW}. In this model, two equivalent parameters 
$a_1=c_1+c_2/N_c$ and $a_2=c_2+c_1/N_c$
describe the external and internal $W$-emission amplitudes, respectively. 
However, the evaluation of the  hadronic form factors usually involve 
some ansatz \cite{CT} and thus 
the extraction of $a_1$ and $a_2$ is model dependent. 
It is also found that only when the experimental errors are expanded greatly 
does an allowed domain 
$(a_1,a_2)$ exist for the three classes of decays ${\bar B}^0\to D^{(*)+}$, 
${\bar B}^0\to D^{(*)0}$ and $B^-\to D^{(*)0}$ \cite{GKKP}.
In addition, a negative $a_2/a_1$ and a positive $a_2/a_1$ are concluded 
from the data of charm and bottom decays \cite{BSW,KP}, respectively.

It was shown recently that the perturbative QCD (PQCD) approach 
is applicable to heavy meson decays \cite{L1,WYL}, to which the heavy 
mass provides the large momentum transfer. The breakthrough 
is due to the all-order Sudakov resummation of large radiative 
corrections, which suppress contributions from the long-distance region. 
This formalism, taking into account the evolution from the typical scale 
of the hard subprocesses characterized by the heavy meson mass to a lower 
hadronic scale, is $\mu$ independent for semileptonic decays. 
In this letter we shall develop a PQCD formalism based on the effective 
Hamiltonian in Eq.~(\ref{eff}), which further incorporates the evolution 
from $M_W$ down to the hard scale.
This three-scale factorization theorem, being $\mu$ independent, does
not suffer the scale-setting ambiguity mentioned above. We apply this
formalism to two-body nonleptonic $B$ meson decays such as $B\to D\pi$. 
Without any free parameter, our prediction agrees well with experimental 
data.

We first illustrate the main idea of PQCD factorization theorems by 
considering one-loop QCD corrections to a generic decay process through
a current. These corrections are ultraviolet finite, since the conserved 
current is not renormalized. However, they also give rise to 
infrared divergences, when the gluons are soft or collinear to light 
partons. The factorization is implemented to isolate these infrared 
divergences associated with the long-distance physics. 

Radiative corrections that produce infrared divergences are classified into 
the reducible and irreducible types \cite{LS}. Irreducible corrections 
contain only single soft logarithms and is absorbed into a soft function
$U$, which corresponds to the nonfactorizable soft corrections in $B$ meson 
decays in the literature \cite{CT}. These corrections cancel asymptotically 
\cite{LS}, 
and thus are expected to be small. They will be neglected here 
({\it ie} $U=1$) and studied in a forthcoming work. Reducible corrections 
contain double logarithms from the combination of soft and collinear 
divergences, which can be absorbed into a wave function $\phi(P,b,\mu)$ and 
explicitly resummed into a Sudakov factor \cite{LS}, 
\begin{equation}
\phi(P,b,\mu)=\exp[-s(P,b)]\phi(b,\mu)\;.
\label{s}
\end{equation}
$b$ is the conjugate variable of the transverse momentum, which will be 
explained later, and $1/b$ can be regarded as an infrared cutoff. 

To factorize a one-loop correction, we divide it into two terms as shown 
in Fig.~1(a). The first term, with eikonal approximation for fermion 
propagators, picks up the infrared structure of the full diagram. 
Being infrared sensitive, it is absorbed into $U$ or $\phi$, 
depending on which type the one-loop correction is.
The second term, with a soft subtraction, is infrared
safe. It has the same ultraviolet structure as the full diagram and 
can be absorbed into a hard scattering amplitude $H(t,\mu)$, where $t$ 
denotes the typical scale of the hard decay process. We then get the 
$O(\alpha_s)$ factorization formula shown in Fig.~1(b), where the 
diagrams in the first parentheses contribute to $H$.

The presence of $\mu$ implies that both $\phi$ and $H$ need renormalization. 
Let $\gamma_\phi$ be the anomalous dimension of $\phi$. Then the anomalous 
dimension of $H$ must be $-\gamma_\phi$, because the full diagram does not 
contain ultraviolet divergences. Their $\mu$ dependence can be calculated  
by RG,
\begin{eqnarray}
\phi(b,\mu)&=&\phi(b,1/b)\exp\left[-\int_{1/b}^\mu\frac{d{\bar\mu}}
{\bar\mu}\gamma_\phi(\alpha_s({\bar\mu}))\right] \;,
\label{phi} \\
H(t,\mu)&=&H(t,t)\exp\left[-\int_{\mu}^t\frac{d{\bar\mu}}
{\bar\mu}\gamma_\phi(\alpha_s({\bar\mu}))\right]\;.
\label{h}
\end{eqnarray}
Equation (\ref{phi}) describes the evolution of $\phi$ from $1/b$ to an 
arbitrary scale $\mu$, and (\ref{h}) describes the evolution of $H$ from 
$\mu$ to $t$. The contribution characterized by momenta smaller than $1/b$, 
{\it ie.}, the infrared divergence, is absorbed into the initial condition 
$\phi(b,1/b)$, which is of nonperturbative origin.
The convolution of $H$ with $\phi$ is then $\mu$ independent as indicated by
\begin{equation}
H(t,\mu)\phi(b,\mu)=H(t,t)\phi(b,1/b)
\exp\left[-\int_{1/b}^t\frac{d{\bar\mu}}
{\bar\mu}\gamma_\phi(\alpha_s({\bar\mu}))\right]\;.
\label{phih}
\end{equation}
In this way all the large single logarithms are collected in the 
exponential. 

Indeed the effective Hamiltonian in Eq.~(\ref{eff}) can be constructed in 
a similar way. Consider now a typical one-loop QCD correction to the $W$ 
boson exchange diagram Fig.~1(c). We express the full diagram, which is 
ultraviolet 
finite, into two terms as shown in Fig.~1(c). The first term, obtained 
by shrinking the $W$ boson line into a vertex, corresponds to the local 
four-fermion operators $O_i$. It is absorbed into a hard scattering 
amplitude $H(t,\mu)$, with a typical scale $t\ll M_W$, since gluons 
involved in this term do not ``see" the $W$ boson. Its dependence on 
$M_W$ is limited to the $1/M_W^2$ factor of the four-fermion operators. 
The second term, characterized by momenta of order $M_W$,
is absorbed into a ``harder" function $H_r(M_W,\mu)$ (not a scattering 
amplitude), in which gluons do ``see" the $W$ boson. 

We obtain the $O(\alpha_s)$ factorization formula shown 
in Fig.~1(d), where the diagrams in the first parentheses contribute to 
$H_r$, and those in the second parentheses to $H$. Note that this formula 
in fact represents a matrix relation because of the mixing between 
operators $O_1$ and $O_2$. Solving their RG equations, we derive
\begin{equation}
H_r(M_W,\mu)H(t,\mu)=H_r(M_W,M_W)H(t,t)
\exp\left[\int_{t}^{M_W}\frac{d{\bar\mu}}
{\bar\mu}\gamma_{H_r}(\alpha_s({\bar\mu}))\right]\;,
\label{hrh}
\end{equation}
where the anomalous dimension $\gamma_{H_r}$ of $H_r$ is also a matrix. 
We emphasize that the factorization in Eq.~(\ref{hrh}) is not complete
because of the presence of infrared divergences in $H$.
The exponential can be easily identified as the Wilson coefficient, 
implying that $\mu$ in $c(\mu)$ should be set to the hard scale $t$.
Without large logarithms, $H_r(M_W,M_W)$ can now be safely approximated 
by its lowest-order expression $H^{(0)}_r=1$. 

We are now ready to contruct a three-scale factorization theorem by 
combining Eqs.~(\ref{phih}) and (\ref{hrh}). Consider the decay 
amplitude up to $O(\alpha_s)$ without integrating out the $W$ boson. We 
first factorize out the infrared sensitive wave functions as described above.
Though devoid of infrared divergences (the nonfactorizable soft corrections
have been neglectd here), the hard part still invloves 
two scales $t$ and $M_W$. The factorization in Fig.~1(d) is then employed 
to separate these two scales, and $H_r$ can be moved out of the hard part, 
a step valid up to $O(\alpha_s)$. 
We identify the remaining diagrams, including the four-fermion amplitude 
and the associated soft subtraction, as the hard scattering amplitude $H$, 
since it is free of infrared divergences. The anomalous dimension of $H$
is given by $\gamma_H=-(\gamma_\phi+\gamma_{H_r})$. We thus 
get the three-scale factorization formula
\begin{eqnarray}
H_r(M_W,\mu)H(t,\mu)\phi(b,\mu)=c(t)H(t,t)\phi(b,1/b)
\exp\left[-\int_{1/b}^t\frac{d{\bar\mu}}
{\bar\mu}\gamma_\phi(\alpha_s({\bar\mu}))\right]\;,
\label{main}
\end{eqnarray}
with the Wilson coefficient $c(t)$ given by the exponential factor in 
Eq.~(\ref{hrh}). The two-stage evolutions from $1/b$ to $t$ and from $t$ 
to $M_W$ are both included, and the final expression is $\mu$ independent.

The above conclusion is quite natural from the effective field theory 
approach \cite{Ge}. An effective field theory is constructed for a 
scale $\mu < M_{W}$ by integrating out the $W$ boson at $\mu=M_{W}$. 
Matching corrections are determined by the matching condition
requiring that the low-energy light-particle Green functions of the 
two theories be equal. 
The effective theory is then evolved by RG running from 
$\mu=M_{W}$ to a lower scale, which insures that the amplitudes
are $\mu$ independent.
The scale $\mu$ in a continuum effective field theory is 
actually a scale to separate the long-distance from 
the short-distance physics with the physics above the scale 
$\mu$ absorbed into the coefficients in the effective Hamiltonian, such 
as the Wilson coefficients $c_{1,2}(\mu)$. 
This idea is identical to the PQCD factorization theorem.
The effective field theory constructed this way has exactly the same 
low-energy behaviour as the full theory, including infrared divergences, 
physical cuts, and etc. Thus the infrared divergences in the decay 
amplitudes calculated using the effective field theory can be factorized 
in the same way as the full theory. The factorization formula for the
$\mu$ independent amplitude is identical to Eq.~(\ref{main}), 
\begin{equation} 
c(M_W,\mu) H(t,\mu)\phi(b,\mu)\;,
\end{equation}
with the Wilson coefficient $c$ identified as $H_r$.

We now apply the above formalism to the nonleptonic decays 
$B(P_{1}) \to D(P_{2}) \, \pi(P_{3})$. The decay rate can 
be written as
\begin{equation}
\Gamma=\frac{1}{128\pi}G_F^2|V_{cb}|^2|V_{ud}|^2M_B^3\frac{(1-r^2)^3}{r}
|{\cal M}|^2\;,
\end{equation}
with $r=M_{D}/M_B$, $M_B$ ($M_D$) being the $B$ ($D$) meson mass. In the 
rest frame of the $B$ meson, $P_1$ has 
the components $P_1=(M_B/\sqrt{2})(1,1,{\bf 0}_T)$. The nonvanishing 
components of $P_2$ and $P_{3}$ are respectively
$P_2^+=M_{B}/\sqrt{2}$, $P_2^-=rM_{D}/\sqrt{2}$, $P_3^+= 0$, and
$P_3^-=(1-r^2)M_{B}/\sqrt{2}$. 
Let $k_{1}$($k_{2}$) be the momentum of the light valence quark in the $B$ 
($D$) meson and $k_{3}$ be the momentum of a valence quark in the pion. 
These $k$'s may be off-shell by the amount of their transverse components
$k_T$ of order $\Lambda_{\rm QCD}$. We define the momentum fractions $x$ as 
$x_{1}=k_{1}^{-}/P_{1}^{-}$, $x_{2}=k_{2}^{+}/P_{2}^{+}$, and
$x_{3}=k_{3}^{-}/P_{3}^{-}$.
The transverse momenta $k_{iT}$ play the role of an infrared cutoff in
our analysis.

To leading power in $1/M_{B}$, the factorization formula for ${\cal M}$ 
in the transverse configuration space \cite{LS} is written as 
\begin{eqnarray}
{\cal M}&=& \int_{0}^{1} [dx]\int_{0}^{\infty} [d^2{\bf b}]
          \phi_{B}(x_{1},{b}_{1},1/b_1) 
          \phi_{D}(x_{2},{b}_{2},1/b_2) 
          \phi_{\pi}(x_{3},{b}_{3},1/b_3)\,  \nonumber   \\
      & & \times  c(t) H(x_i,b_i,t)\exp[-S(x_i,b_i)] \,
\end{eqnarray}
with $[dx]=dx_{1}dx_{2}dx_{3}$ and $[d^2{\bf b}]=d^2{\bf b}_{1}
d^2{\bf b}_{2}d^2{\bf b}_{3}$. 
The Sudakov factor 
$e^{-S}$ is the product of $e^{-s}$ in Eq.~(\ref{s}) and the exponential 
in Eq.~(\ref{main}) from each wave function. In the analysis below we 
shall neglect the $b$ dependence of the wave functions \cite{L1}.

Without large logarithms, the hard part $H$ can be reliably treated by 
perturbation theory. To leading order in $\alpha_{s}$, the hard part 
for the decay $B^-\to D^0\pi^-$ consists 
of four sets of diagrams shown in Fig.~2. The diagrams in Fig.~2(a) 
correspond to the external $W$ emission \cite{BSW,NRSX}, while 
those in Fig.~2(b) to the internal $W$ emission. They have been calculated 
using the PQCD formalism in \cite{L1,WYL} without including the Wilson 
coefficients. Denote their contributions to the amplitude $\cal M$ as 
${\cal M}_{a}$ and ${\cal M}_{b}$. It is easy to find that the Wilson 
coefficients associated with ${\cal M}_{a}$ and ${\cal M}_{b}$ are 
respectively $a_1$ and $a_2$. Readers are refered to 
\cite{WYL} for the complete formulas of ${\cal M}_{a}$ and ${\cal M}_b$.

Diagrams in Fig.~2(c) and 2(d) are absent in the factorization 
approximation and will be called the nonfactorizable diagrams.
Fig.~2(c) leads to the amplitude ${\cal M}_c$
\begin{eqnarray}
 {\cal M}_{c}&=& 32 \sqrt{2N_{c}} \pi{\cal C}_F\sqrt{r}M_B^2 G_{F}
\int_0^1 [dx]\int_0^{\infty}b_1 db_1 b_2 db_2
\phi_B(x_1)\phi_D(x_2)\phi_\pi(x_3) \nonumber \\
& &\times \left[ \alpha_s(t_1)  
\frac{c_{1}(t_1)}{N_{c}}  e^{-S_{c1}(x_i,b_i)}
(x_1-x_2-x_3(1-r^2))h^{(1)}_{c}(x_i,b_i) \right.
\nonumber \\
& &  \left.  + \alpha_s(t_2)  
\frac{c_{1}(t_2)}{N_{c}}  e^{-S_{c2}(x_i,b_i)}
(1-(x_1+x_2)(1-r^2))h^{(2)}_{c}(x_i,b_i) \right]\;.
\end{eqnarray}
The functions $h_c^{(j)}$, $j=1$ and 2, are given by
\begin{eqnarray}
\everymath{\displaystyle}
h^{(j)}_{c}&=& \left[\theta(b_1-b_2)K_0\left(AM_B
b_1\right)I_0\left(AM_Bb_2\right)
+\theta(b_2-b_1)K_0\left(AM_B b_2\right)
I_0\left(AM_B b_1\right)\right]\;  \nonumber \\
&  & \times \left( \begin{array}{cc}
 K_{0}(B_{j}M_Bb_{2}) &  \mbox{for $B_{j} \geq 0$}  \\
 \frac{i\pi}{2} H_{0}^{(1)}(|B_{j}|M_Bb_{2})  & \mbox{for $B_{j} \leq 0$}
  \end{array} \right)\;,           
\end{eqnarray}
with $A^{2}=x_{1}x_{3}(1-r^{2})$,
$B_{1}^{2}=(x_{1}+x_{2})r^{2}-(1-x_{1}-x_{2})x_{3}(1-r^{2})$, and
$B_{2}^{2}=(x_{1}-x_{2})x_{3}(1-r^{2})$.
The Sudakov exponent $S_{cj}$ is written as 
\begin{eqnarray}
S_{cj}&=&s(x_1P_1^+,b_1)+s(x_2P_2^+,b_2)+
s((1-x_2)P_2^+,b_2) + s(x_3P_3^-,b_3)
\nonumber \\
& &+s((1-x_3)P_3^-,b_3)
-\frac{1}{\beta_1}\sum_{i=1}^{3}\ln\frac{\ln(t_j/\Lambda)}
{-\ln(b_i\Lambda)}\;,
\end{eqnarray}
with $b_3=b_2$, $\beta_1=(33-2n_f)/12$ and $n_f=4$ the number of flavors. 
The scale $t_j$ is chosen as $t_j={\rm max}(AM_B,|B_j|M_B,1/b_1,1/b_2)$, 
and $\Lambda\equiv\Lambda_{\rm QCD}$ is set to 0.2 GeV. 
The Sudakov suppression described by the factor 
$\exp(-S_c)$ then warrants that main contributions come from 
the small $b$, or large $t$, region, in which $\alpha_s(t)$ is small, and 
thus the perturbative treatment of the hard part is reliable.
The amplitude ${\cal M}_d$ is obtained from Fig.~2(d) accordingly.
The amplitudes for the decay ${\bar B}^0\to D^+\pi^-$ can be derived 
in a similar way. However, it is found that only the external
$W$-emission contribution, the same as ${\cal M}_a$, is important.
Therefore, we shall not give its expression here.

The wave functions are chosen as
\cite{WYL},
\begin{equation}
\phi_\pi(x)=\frac{5\sqrt{6}}{2}f_\pi x(1-x)(1-2x)^2
\;,\;\;\;\;
\phi_{B,D}(x)=\frac{N_{B,D}}{16\pi^2}\frac{x(1-x)^2}
{M_{B,D}^2+C_{B,D}(1-x)}\;,
\end{equation}
where $f_\pi=132$ MeV is the pion decay constant.
$N_B=650.212$ and $C_B=-27.1051$ correspond to the $B$ meson
decay constant $f_B=200$ MeV. $N_D$ is determined by the $D$ meson 
decay constant $f_D=220$ MeV, and $C_D$ is
fixed by data for the decay ${\bar B}^0\to D^+\pi^-$ \cite{WYL}.
All other parameters are referred to \cite{WYL}.

The experimental data of the branching ratios are 
${\cal B}_0={\cal B}({\bar B}^0 \to D^{+} \pi^{-})=(3.08\pm 0.85) 
\times 10^{-3}$ and ${\cal B}_-={\cal B}(B^{-} \to D^{0} \pi^{-})
=(5.34\pm 1.05) \times 10^{-3}$ 
\cite{A}. Our predictions using the original Hamiltonian in 
Eq.~(\ref{FourFer}), {\it ie} $c_1=1$ and $c_2=0$, are 
${\cal B}_0=3.08\times 10^{-3}$ and ${\cal B}_-=5.10\times 10^{-3}$. 
If the three-scale factorization formulas based on the effective 
Hamiltonian is employed, we obtain ${\cal B}_0=3.08\times 10^{-3}$ and
${\cal B}_-=5.00\times 10^{-3}$, which differ from the previous 
results only by 2\%. The slight variation justifies the analysis of
exclusive nonleptonic $B$ meson decays performed in \cite{WYL} using 
Eq.~(\ref{FourFer}). A careful observation reveals that when the evolution 
of the Wilson coefficients is taken into account, the amplitude 
${\cal M}_b$, which is proportional to $a_2(t)$, becomes smaller, while 
${\cal M}_c$, which is
proportional to $c_1(t)/N_c$, becomes larger. Hence, the two changes 
cancel each other, and the total decay rate remains almost the same. 
${\cal M}_d$ is less important because of the pair cancellation between
the two diagrams in Fig.~2(d). Our calculation also indicates that 
the nonfactorizable contribution ${\cal M}_{c}$ 
is substantial, and the limit of the BSW factorization 
approximation. That is why the naive choice of $a_{1,2}=a_{1,2}(M_B)$ in 
the BSW model fails to explain the data. 

Applying the three-scale factorization theorem to
the mode $D^-\to K^0\pi^-$, we obtain the predictions 
${\cal B}({\bar D}^0\to K^+\pi^-)=4.05\%$ and
${\cal B}(D^-\to K^0\pi^-)=2.67\%$, consistent with the data
$(4.01\pm 0.14)\%$ and $(2.74\pm 0.29)\%$, respectively.
With the running scale $t$ reaching below the $c$ quark mass, ${\cal M}_b$  
becomes more negative and overcomes 
the positive contribution of ${\cal M}_c$. This explains the 
observed destructive interference of the external and internal W-meson 
emission contributions absent in the $B$ meson decays. 
Hence, the nonfactorizable diagrams play an important role in the
explanation of the charm decay data.
It is clear that a PQCD formalism  based on the original 
Hamiltonian without Wilson coefficients \cite{WYL,CM} 
can not account for this change of sign in the charm 
decays. From the above analysis, we suggest that $c_{1,2}$ could be 
regarded as Wilson coefficients as they originally are, if the scale is 
chosen properly, instead of as fitting parameters in the BSW model.
That is, the controversy over the extraction of $a_2/a_1$ from the bottom  
and charm decays does not exist in our theory.

The scale dependence of our formalism can be tested by substituting
$2t$ for $t$ in the factorization formula. It is found that the 
prediction decreases by only 5\%. In the conventional effective
field theory, the substitution of $M_b$ by $2M_b$ for the argument of
the Wilson coefficients $c_{1,2}$ results in a 10\% to 20\% difference
\cite{LSW}. Hence, the scale-setting ambiguity is moderated in the 
three-scale factorization theorems. In conclusion, our formalism provides 
a more sophisticated choice of the scale, which takes into account the 
additional low-energy dynamics in the mesons. This approach 
is expected to give more definitive predictions, when it is applied to 
inclusive nonleptonic $B$ meson decays. This subject will be discussed
elsewhere. 

Our formalism can also be applied to the decays $B\to J/\Psi K^{(*)}$.
With the three-scale factorization formulas and the 
inclusion of nonfactorizable contributions,
both the decay rates and the fraction of the longitudinal mode can
be explained. The details will be published in a separate work.

\vskip 0.5cm
This work was supported by the National Science Council of ROC under
Grant No. NSC-85-2112-M-194-009 (for H. L.) and NSC-85-2112-M007-029, 
NSC-85-2112-M007-032 (for C. C.).

\newpage
\centerline{\large \bf Figure Captions}
\vskip 0.5cm

\noindent
{\bf Fig. 1.} (a) Separation of infrared and hard $O(\alpha_s)$ 
contributions in PQCD. (b) $O(\alpha_s)$ factorization into a wave 
function and a hard scattering amplitude. (c) Separation of hard and harder 
$O(\alpha_s)$ contributions in an effective field theory. (d) $O(\alpha_s)$ 
factorization into a ``harder" function and a hard scattering 
amplitude.
\vskip 0.5cm

\noindent
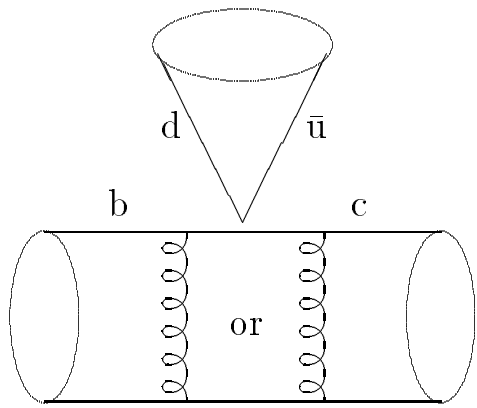
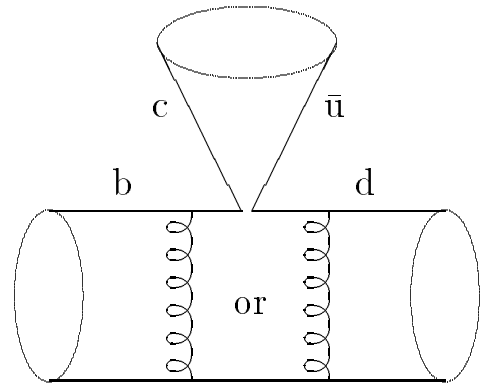
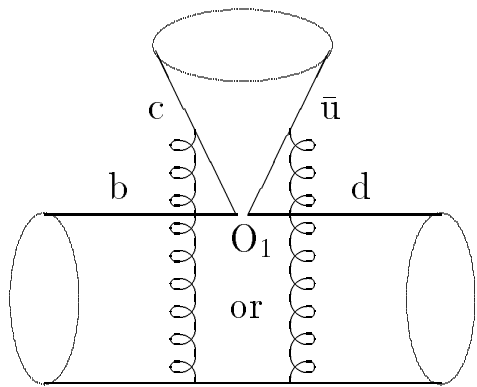
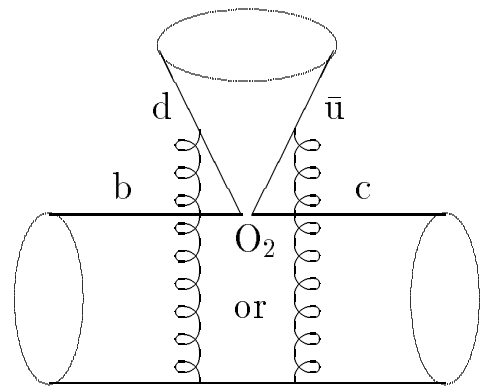
{\bf Fig. 2.} (a) External $W$ emission. (b) Internal $W$ emission.
(c) and (d) Nonfactorizable internal $W$ emission. 

\end{document}